\documentclass{article}

\usepackage{arxiv}

\usepackage[utf8]{inputenc} 
\usepackage[T1]{fontenc}    
\usepackage{hyperref}       
\usepackage{url}            
\usepackage{booktabs}       
\usepackage{amsfonts}       
\usepackage{nicefrac}       
\usepackage{microtype}      
\usepackage{lipsum}		
\usepackage{graphicx}
\usepackage{natbib}
\usepackage{doi}
\usepackage{amsmath}
\usepackage{algorithm}
\usepackage{algorithmic}
\usepackage{multirow}
\usepackage{makecell}

\title{PI-AstroDeconv: A Physics-Informed Unsupervised Learning Method for Astronomical Image Deconvolution}

 \author{Shulei Ni, Yisheng Qiu, Yunchun Chen, Zihao Song, Hao Chen, Xuejian Jiang, and Huaxi Chen \thanks{Corresponding author}\\
      Research Center for Astronomical Computing,\\
      Zhejiang Laboratory, 311121 Hangzhou, Zhejiang, China \\
    \texttt{\{nisl, chenych, songzihao, jiangxuejian, chenhuaxi\}@zhejianglab.com}\\
    \texttt{yishengq@126.com, haochen.cluster@gmail.com}}

\renewcommand{\shorttitle}{\textit{arXiv} Template}

\hypersetup{
pdftitle={A template for the arxiv style},
pdfsubject={q-bio.NC, q-bio.QM},
pdfauthor={David S.~Hippocampus, Elias D.~Striatum},
pdfkeywords={First keyword, Second keyword, More},
}

\begin{document}
\maketitle

\begin{abstract}
	In the imaging process of an astronomical telescope, the deconvolution of its beam or Point Spread Function (PSF) is a crucial task. However, deconvolution presents a classical and challenging inverse computation problem. In scenarios where the beam or PSF is complex or inaccurately measured, such as in interferometric arrays and certain radio telescopes, the resultant blurry images are often challenging to interpret visually or analyze using traditional physical detection methods. We argue that traditional methods frequently lack specific prior knowledge, thereby leading to suboptimal performance. To address this issue and achieve image deconvolution and reconstruction, we propose an unsupervised network architecture that incorporates prior physical information. The network adopts an encoder-decoder structure while leveraging the telescope's PSF as prior knowledge. During network training, we introduced accelerated Fast Fourier Transform (FFT) convolution to enable efficient processing of high-resolution input images and PSFs. We explored various classic regression networks, including autoencoder (AE) and U-Net, and conducted a comprehensive performance evaluation through comparative analysis. 

\end{abstract}

\section{Introduction}

Astronomical image deconvolution is a complex and inherently ambiguous problem in the domain of astronomical observation. It has been extensively studied for decades and is considered a classic issue in inverse computational imaging. Moreover, it has gained considerable attention within the field of image processing. In the context of radio astronomy, eliminating the beam effects generated by the telescope is of utmost importance for achieving precise and accurate images~\citep{beckers1994effects,mort2016analysing,Ni:2022kxn}.

Sidelobes are commonly found in the PSF or Beam of almost all telescopes, especially those based on interferometry~\citep{covington1957interferometer, woody2001radio, woody2001radio2}. These sidelobes emerge as a result of factors like optical aberrations, diffraction, and imperfections in the imaging system. Sidelobes can pose challenges in tasks related to image processing and analysis, as they have the potential to introduce undesired artifacts and impact the overall quality of the image. In the case of deconvolution algorithms, for instance, sidelobes can be mistakenly identified as genuine image features during the reconstruction process, leading to errors and a loss of detail~\citep{jackson2008principles,tsao1988reduction}.

The inherent beam effects in astronomical telescopes inevitably distort the measured data or obtained images of observed objects. These effects can arise from imperfections in the telescope's imaging system or atmospheric conditions. The presence of beam effects can lead to image blurring or spatial distortions, significantly impacting image clarity and resolution. Given that a single input image may correspond to multiple potential clear images, the field of image deconvolution becomes inherently challenging and problematic~\citep{beckers1994effects,mort2016analysing,Rohlfs1996}. Consequently, it becomes increasingly vital to effectively eliminate these effects, enhance image quality, and enable scientists to conduct detailed studies of the observed objects.

It is important to acknowledge that sidelobes cannot be entirely eliminated, as they are inherent to the imaging system and the physics of the image formation process. Nonetheless, by meticulous algorithm design and calibration, the influence of sidelobes can be substantially mitigated, resulting in enhanced image quality and more precise analysis outcomes.

When tackling the deconvolution problem, conventional algorithms (e.g., ~\citet{kundur1996blind,krishnan2009fast,pan2014motion,ren2016image,pan2016l_0}) typically seek to locate the optimal solution through the inference of the convolution kernel.
Deconvolving images primarily involves solving a highly nonlinear and uncertain optimization problem, which makes successful deconvolution extraordinarily challenging. The deconvolution task becomes even more difficult if the provided convolution kernel is complex or if the measurements are imprecise. Some deconvolution methods attempt to overcome convolution by incorporating various image priors, such as the red-dark channel prior~\citep{cheng2015underwater,pan2017deblurring} and the gradient prior~\citep{xu2021fast}. However, these methods have limited capabilities in accurately modeling clear image features and generating artifact-free outputs. With the rapid development and application of deep learning methods in the field of astronomy, they have become more effective approaches for solving such inverse problems due to their ability to handle nonlinearity and large amounts of data. For instance, convolutional neural networks (CNN) have been extensively studied for image deblurring~\citep{xu2014deep,schuler2013machine,nah2017deep,tao2018scale}, with CNN-based methods being widely explored. Among various works, the study by \citet{xu2014deep} has established a connection between traditional optimization-based approaches and neural network structures. Additionally, a new separable structure has been introduced as a dependable means of supporting robust artifact deconvolution.
Subsequently, a supervised network is constructed. Although these methods have produced impressive results~\citep{xu2014deep,dong2020deep,yanny2022deep}, they heavily rely on supervised datasets and deeper, broader architectures to enhance performance. Consequently, deploying them in practical applications presents certain challenges.

In this study we proposed PI-AstroDeconv, a physics-informed unsupervised learning method for astronomical image deconvolution. We utilize observational data\footnote{\url{https://webbtelescope.org/images}} and train the PI-AstroDeconv network using the corresponding PSF simulated by the Webb instrument\footnote{\url{https://github.com/spacetelescope/webbpsf/tree/stable}}. 
This paper consists of the following sections. \ref{sec:ralate_work} presents a comprehensive discussion on the research pertaining to deconvolution. In \ref{sec:our_method}, we introduce our method framework, which encompasses FFT acceleration training and the selection of appropriate loss functions. In \ref{sec:experiments}, we conducted experiments using the established structure, subsequently analyzing and discussing the obtained results. The \ref{sec:conclusion} provides the concluding remarks.

\section{Related Works}\label{sec:ralate_work}
In data processing for radio interferometric arrays and telescopes, CLEAN method is a widely utilized technique for enhancing the quality of single radio interferometric images. The dirty map serves as an initial approximation, albeit with limitations. This approach employs iterative processes to eliminate both the artifacts in the dirty images and the associated beam distortions~\cite{offringa-wsclean-2014,bean2022casa}. Consequently, it does not yield an unique and stable solution, and it demands substantial computational resources. Furthermore, the Maximum Entropy Method (MEM) algorithm is one of the commonly used algorithms, which has advantages in handling extended images~\citep{rohlfs2013tools}. However, both of these algorithms face the challenges of non-uniqueness in solutions and high computational requirements. Therefore, to address these difficulties, we attempt to adopt deep learning algorithms.

Blind deconvolution in astronomical imaging involves two concepts: pure blind deconvolution and physics-informed deconvolution. Pure blind deconvolution refers to the case where neither the convolution kernel nor the input is known. In actual observations, the PSF or Beam of a telescope can be measured. Therefore, we can perform deconvolution based on this physical information. 

The most common and fundamental method for deconvolution is Wiener filtering~\citep{treitel1969predictive}. Wiener filtering is based on frequency domain theory, which can restore the blurreded image and provide local signal enhancement. Its core idea is the minimum mean square error (MMSE) criterion. It assumes that the input image undergoes convolution by a linear time-invariant system and is blurred by the addition of noise. It should be noted that Wiener filtering is a linear and time-invariant filtering method, suitable for deconvolution of images that undergo linear convolution and are affected by Gaussian noise.

\citet{Bai8488519} proposes an image-based blind deconvolution algorithm. The algorithm designs an efficient method for alternatingly solving the skeleton image and the PSF. The authors also propose a reweighted graph total variation (RGTV) prior, which helps in the distribution of bimodal edge weights in the image. Through analysis in the node domain and graph frequency domain, the RGTV prior demonstrates ideal characteristics, such as improved robustness, powerful fragment smoothing filtering, and enhanced image sharpness.

\citet{Chen9578235} explores the challenge of blind deblurring for overexposed images. It emphasizes that conventional methods frequently struggle with restoring clarity in overexposed images due to the non-compliance of pixels surrounding saturated areas with the commonly adopted linear blur model. To tackle this issue, the authors introduce a novel blur model that accommodates both saturated and unsaturated pixels, enabling the inclusion of all informative pixels during the deblurring process.

\citet{Nan9157433} discusses the problem of kernel/model error in non-blind image deconvolution methods. It proposes a deep learning approach that takes into account the uncertainty of blurry kernel and convolution models. This approach utilizes a total least squares estimator and priors learned from neural networks to handle the kernel/model error.

\citet{Ren9156633} presents a novel neural optimization solution for the problem of blind deconvolution, which is a challenging low-level vision problem. Traditional approaches rely on fixed and handcrafted priors, which are insufficient to describe clean images and convolution kernels. Existing deep learning methods can handle complex convolution kernels, but they have limited scalability. Therefore, the authors propose a generative network called SelfDeblur to simulate priors of clean images and convolution kernels. This network combines asymmetric autoencoders and fully connected networks to respectively generate latent clean images and convolution kernels.

In light of the image generation process of astronomical telescopes and the substantial dimensions of the PSF or Beam, we propose a deconvolution model that relies on physical information. Furthermore, we conducted a comparative analysis of this approach with the aforementioned four methods.

\section{Our Method}\label{sec:our_method}
This section will primarily discuss the network architecture of PI-AstroDeconv, the FFT-accelerated convolution method, and the selection of appropriate loss functions.

\subsection{Overview}

\begin{figure*}
	\centering
	\includegraphics[width=0.9\textwidth]{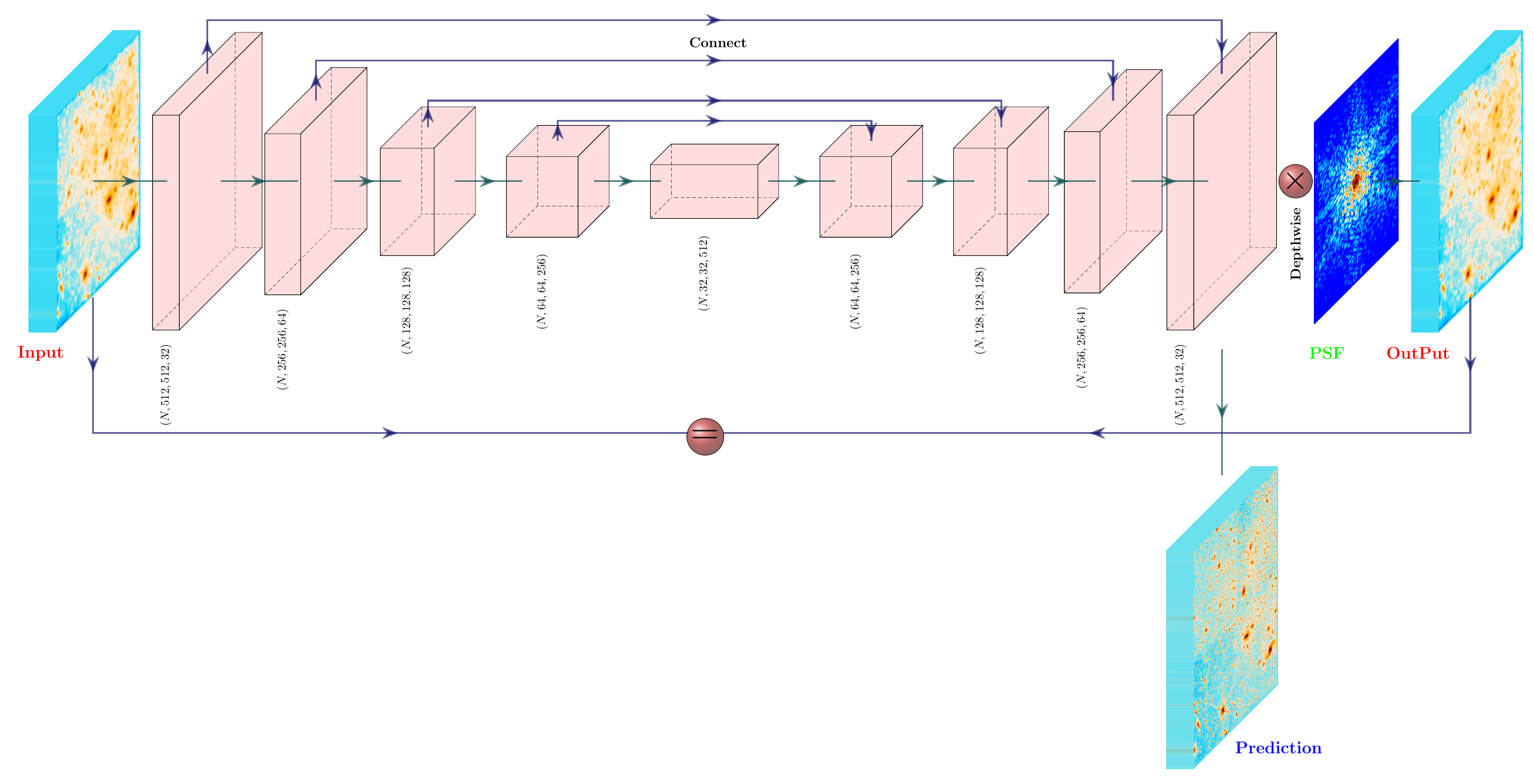}
	\caption{The PI-AstroDeconv architecture incorporates U-Net as its backbone. The left side of the U-Net represents the input, with the right side incorporating the Point Spread Function (PSF) or telescope beam. According to our network architecture design, during the training phase, the input and output of the network are expected to remain consistent, irrespective of the choice of backbone network. After the training is completed, during the inference phase, the backbone network will directly generate the inference results, which are represented as predictions at the bottom of the diagram.}
	\label{fig:deconv_unet}
\end{figure*}

This paragraph provides a comprehensive overview of the deconvolution architecture to elucidate the underlying structure of the network. Our deconvolution architecture is based on common regression network models, such as AE~\citep{rumelhart1986learning}, U-Net~\citep{ronneberger2015u}, Generative Adversarial Networks~(GAN)~\citep{creswell2018generative} etc., and is applied to imaging devices such as astronomical telescopes to eliminate beam or PSF artifacts. \ref{fig:deconv_unet} shows the architecture of the network. The pink blocks represent the classic U-Net network, with each block representing a network layer rather than a single convolution or pooling operation. The left half of the network is the downsampling path, and the right half is the upsampling path. The box at the bottom indicates the number of channels and output size. The arrows above the diagram represent skip connections in the U-Net. The blocks on the far left and far right represent the input and output of the network, respectively. A deep convolution layer, with a fixed convolution kernel equal to the PSF or Beam of the telescope, is added between the U-Net and the output. The final deep convolution is not a simple reverse multiplication but a mathematically rigorous convolution. Due to the specificity of the network structure, it can be observed that the input and labels (Ground Truth) of the network are the same, which are the observation data from the telescope. The learning objective of the network is to output the same as the input. The final prediction of the network is the last layer of the U-Net, the last pink layer, which corresponds to the desired deconvolved image. This visualization was created using the modified PlotNeuralNet library\footnote{\url{https://github.com/HarisIqbal88/PlotNeuralNet}}. The provided diagram is based on U-Net as an example, but it should be emphasized that other network models mentioned earlier can be used for network training in this architecture.

\subsection{Physics-informed Network}

In order to more effectively eliminate the effects of beam distortion, we have devised an unsupervised image deconvolution framework. This model is grounded on an encoder-decoder network structure, incorporating existing prior knowledge to further enhance performance. The encoder processes blurred images and generates latent clear images through the decoder, which are subsequently convolved with the beam or PSF. This integration guides the learning process and ensures compliance with desired physical properties. Notably, we have observed cases where the beam or PSF size matches the image size of certain telescopes or imaging devices. To preserve the large-scale blurring effect induced by convolutions, the entire image is chosen as the training set, optimizing the utilization of valuable features at various scales in image deblurring.

As illustrated in \ref{fig:deconv_unet}, the PSF operation is incorporated into the final layer of the U-Net network, equipping the PI-AstroDeconv architecture with valuable prior knowledge about the image. Consequently, this inclusion enhances the accuracy of image restoration and deconvolution tasks by accounting for the blurring effects~\citep{racine1996telescope,woody2001radio}. The PSF characterizes how a point source in an image is dispersed or blurred by the imaging system. By incorporating the PSF as prior knowledge, the PI-AstroDeconv architecture achieves a better comprehension of and compensation for blurring effects during the restoration process~\citep{dougherty2001point,cornwell2008multiscale}.

The objective of image restoration is to recover the original sharp image from a blurry or degraded version. Deep learning methods can be trained to learn the mapping between the degraded image and its corresponding sharp image~\citep{nah2017deep,zhang2022deep}. Nonetheless, without any prior information, the network may encounter difficulty in distinguishing between different potential sharp images that could generate the same degraded image~\citep{schuler2013machine}. The utilization of the PSF as prior information empowers the deep learning model to generate more precise and visually pleasing outcomes. Throughout the training process, the PSF serves as a regularization term, guiding the network in producing deblurred images that align with the expected blurring effects. This facilitates the model in restoring details, minimizing artifacts, and enhancing image quality.

By performing coordinate transformations and applying an inverse Fourier transform to the astronomical observation image $I'_D(x, y)$, we can derive the final concise distortion formula for the signal strength $I_D(x, y)$ caused by the telescope antenna $P_D(x, y)$.

\begin{equation}\label{equ:FFT}
	I_D(x, y) = P_D(x, y)\otimes I'(x, y),
\end{equation}
where, for simplicity, we can consider $I_D(x, y)$ as the blurreded image, $I'(x, y)$ as the clear image, and $P_D(x, y)$ as the PSF or beam. The PSF or beam is produced by the Fourier transform of a point source in the regions sampled; this is the response of the interferometer system to a point source.

\subsection{Accelerate convolution through Fourier Transform}
In real-world astronomical observations, the dimensions of both the imaging and the telescope PSF are typically large. In order to preserve the originality of the convolved images and PSF, we refrained from applying any segmentation and instead performed direct convolution calculations. However, utilizing a large convolution kernel ($2048\times2048$) in the network's final layer impeded the learning process. To mitigate this issue, we employed a transformation technique that integrates Fourier transform and convolution. 

The Fourier transform is a mathematical technique utilized in signal processing to convert signals from the time domain to the frequency domain~\citep{boashash2015time}. It has extensive applications in astronomy, specifically in the fields of astronomical signal processing and spectrum analysis. Convolution is a fundamental mathematical operation that plays a crucial role in the advancement of artificial intelligence. However, performing convolution calculations directly in the time domain can be computationally expensive when dealing with large input data. In data processing, convolution operations can be employed as an alternative approach~\citep{connes1970astronomical,starck2002deconvolution,bracewell1956strip}. Let us define a time domain signal $f(t)$ and a convolution kernel $g(t)$. The formula is as follows:
\begin{equation}\label{equ:fft}
	f(t)\otimes g(t)=\text{iFFT}\{\text{FFT}[f(t)]\times \text{FFT}[g(t)]\},
\end{equation}
where, $\text{rFFT}$ represents the real-valued fast fourier transform, and $\text{irFFT}$ represents the inverse real-valued fast fourier transform. The convolution operation in the frequency domain can be substituted by multiplication using the FFT algorithm. In \ref{algor:fftconv}, we provide the pseudocode for FFT-accelerated convolution in neural networks. The symbols $\text{Transpose}$ and $\text{Shape}$ are used for axis alignment and shape retrieval, applicable to both TensorFlow~\citep{abadi2016tensorflow} and PyTorch~\cite{NEURIPS2019_9015}. The symbols \text{ComplexNum} indicates whether the number is complex or not. The $layer$ represents the output of the network, which serves as the input for the convolution operation. The $psf$ stands for Point Spread Function, which is the convolution kernel used in the context of a telescope. We conducted a comparison with the conventional convolution operation. The outcomes of this algorithm exhibit consistent results within the margin of error, employing float32 precision and demonstrating only slight discrepancies up to the seventh decimal place.
\begin{center}
\begin{algorithm}
	\caption{FFT-accelerated Convolution}\label{algor:fftconv}
	\begin{algorithmic}
		\REQUIRE $layer, psf$
		\ENSURE $layer\otimes psf$
		\STATE $layer_T \gets \text{Transpose}(layer, [0, 3, 1, 2])$
		\STATE $s \gets \text{Shape}(layer_T)[-2:] + \text{Shape}(psf)[-2:] - 1$\;
		\IF{\text{ComplexNum}($layer$ \AND $psf$) = \TRUE}
		\STATE FFT, iFFT $\gets$ rFFT2D, irFFT2D
		\ELSE 
		\STATE FFT, iFFT $\gets$ FFT2D, iFFT2D
		\ENDIF
		\STATE $s_{prod} \gets \text{FFT}(layer_T, s) \times \text{FFT}(psf, s)$\;
		\STATE $s_{inver} \gets \text{iFFT}(s_{prod})$\;
		\STATE $\text{start} \gets (\text{Shape}(s_{inver}) - \text{Shape}(layer_T)) // 2$\;
		\STATE $\text{end} \gets \text{start} + \text{Shape}(layer_T)$\;
		\STATE $layer_\text{fftconv} \gets s_{i}\text{[..., start[0]:end[0], start[1]:end[1]]}$\;
		\STATE $layer_\text{fftconv} \gets \text{Transpose}(layer_\text{fftconv}, [0, 2, 3, 1])$
	\end{algorithmic}
\end{algorithm}
\end{center}

This principle is commonly referred to as the convolution theorem, which establishes the equivalence between convolution in the time domain and multiplication in the frequency domain. The FFT is an efficient algorithm employed to compute the Discrete Fourier Transform (DFT)~\citep{fialka2006fft,hurchalla2010time,mathieu2013fast}. It capitalizes on the symmetric features of complex exponentials, leading to a considerable reduction in computational complexity when compared to direct DFT calculations. The notable advantage of utilizing FFT for convolution is its ability to decrease the computational complexity from $\mathcal{O}(n^4)$ to $\mathcal{O}(n^2 log n)$, with $n$ representing the input size. 
The solution was examined through tests utilizing our training GPU, namely the NVIDIA A40. While the regular convolution process took $12832.64$ seconds, our approach achieved the same task in a mere 0.68 seconds, resulting in a 10000-fold increase in speed compared to the regular convolution. Consequently, FFT emerge as a more expedient choice, particularly for larger input sizes~\citep{mathieu2013fast,zhang2020fast}.

\subsection{Loss Functions}
Our objective is to perform image deconvolution on the blurred image data using a supervised regression algorithm that predicts continuous output values based on input values. When selecting a loss function, it is essential to ensure the continuity and differentiability of the information. We have explored the use of various regression loss functions, including MAE (L1 norm), MSE (L2 norm), Huber, and Log-Cosh~\citep{wang2020comprehensive}. However, our experiments with SSIM and PSNR as loss functions yielded unsatisfactory results. The discrepancy arises from the fact that in astronomical data, each small image block may correspond to a galaxy, rendering each pixel critical for data analysis. Considering the robustness to outliers and the second-order differentiability properties of the Log-Cosh loss function, we give preference to its implementation. Log-Cosh is a logarithmic hyperbolic cosine loss function that calculates the logarithm of the hyperbolic cosine of the prediction error. The formula is as follows:
\begin{equation}\label{equ:loss_fun}
	L(p, t)=\sum_i\log\cosh(p_i-t_i).
\end{equation}

The Log-Cosh function demonstrates similarities to MAE for small losses and MSE for large losses, and it is second-order differentiable. On the other hand, the Huber loss function lacks differentiability in all cases. MAE loss represents the average of absolute errors and fails to address significant errors in predictions by only considering the average absolute distance between the predicted and expected data. MSE loss, on the other hand, emphasizes significant errors with squared values, which has a relatively large impact on the performance indicator. Hence, we have selected the log-cosh function due to its superior resistance to outliers.

\section{Experiments}\label{sec:experiments}
This chapter primarily presents the experimental data, network parameter settings, and the discussion of results.

\subsection{Datasets and Experimental Setting}


In astronomical observations, the images of celestial objects captured through telescopes are inherently blurred due to the PSF. Telescope images are obtained in a linear manner where the targets are segmented, individually imaged, and subsequently added and combined. 
The PSF affects the appearance of the image by convolving with the celestial image, producing an impact at the pixel level.
Consequently, eliminating the influence of PSF poses a significant challenge. To showcase the efficacy of our network in addressing this issue, we will assess its performance using Webb data, which is the most commonly used astronomical dataset. Moreover, we will strive to establish a robust correspondence between astronomical images and simulated PSF. Subsequently, the images will undergo convolution with PSF to acquire blurred images that will serve as inputs and labels for the PI-AstroDeconv network. It is important to note that even if the image does not perfectly align with the PSF, our approach remains successful as it primarily focuses on the convolution capability of the network.

\begin{figure*}
	\centering
	\includegraphics[width=0.8\textwidth]{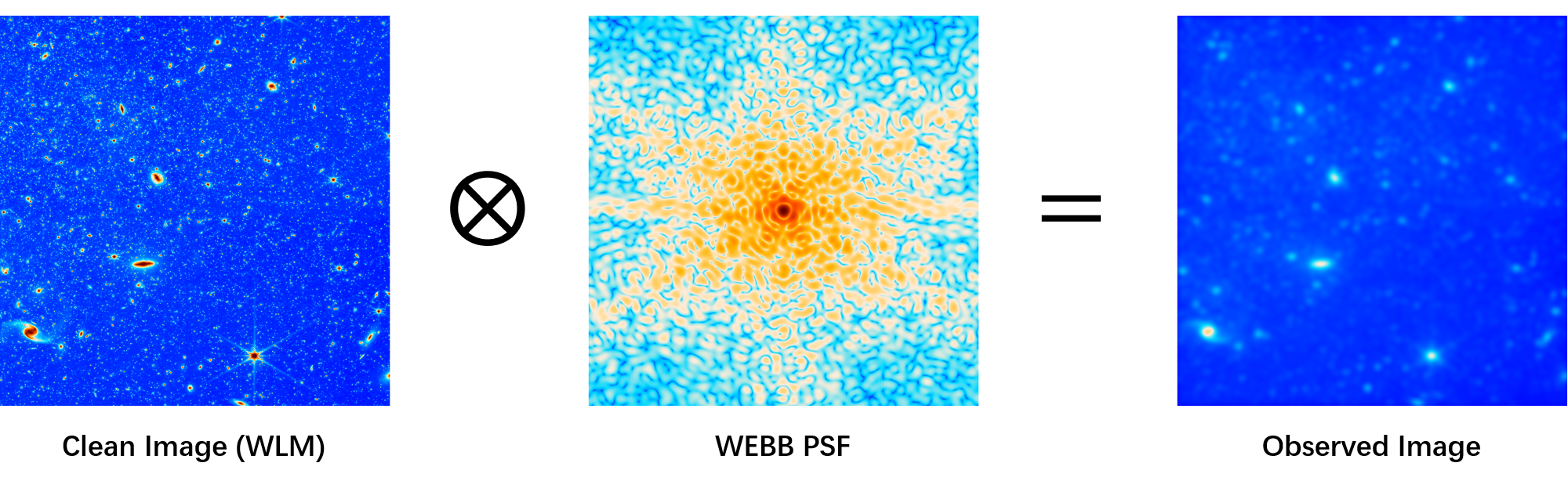}
	\caption{Sketch of telescope observation effects. The image on the left corresponds to the clean image, and the image in the middle showcases the PSF or beam of the telescope. The image on the right illustrates the blurring effect brought by the PSF or beam. The symbol in the middle represents the convolution operation.}
	\label{fig:conv_simp}
\end{figure*}
These images were synthesized from individual exposures captured by the James Webb Space Telescope using the NIRCam instrument~\citep{burriesci2005nircam,horner2004near}. Different filters were employed to capture various infrared wavelength ranges. The colors in the images were obtained by assigning different hues to monochromatic (grayscale) images associated with each filter. We also simulated the corresponding PSF using webbpsf. Considering the actual observational effects, the PSF was convolved with the images, resulting in deconvolved images.

\begin{figure*}
	\centering
	\includegraphics[width=0.9\textwidth]{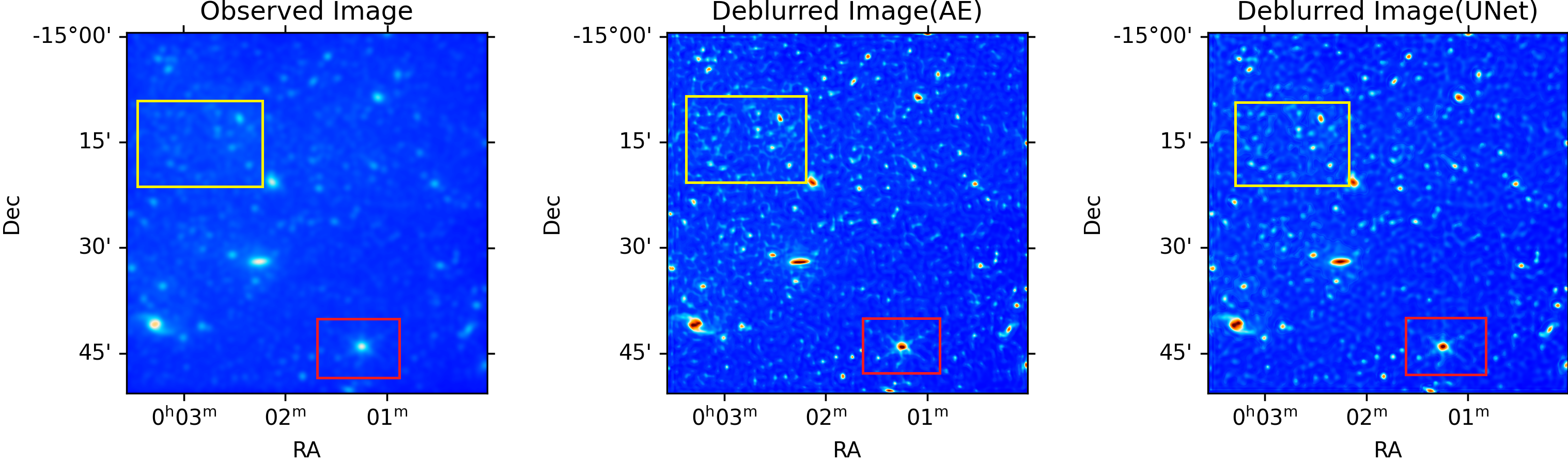}
	\caption{The input/output, and deconvolved images of the PI-AstroDeconv network architecture. The left panel illustrates both the input and output of the PI-AstroDeconv network architecture. In the middle, the image displays the deconvolutional result obtained using the Autoencoder. The right panel showcases the deconvolutional outcomes achieved with UNet. The horizontal and vertical coordinates in the figure represent right ascension and declination, respectively, indicating the corresponding position on the celestial map.}
	\label{fig:image_patch}
\end{figure*}

The image data released by James Webb is synthesized from multiple detectors. For example, the image of dwarf galaxy Wolf–Lundmark–Melotte~(WLM) observed by NIRCam is composed of four filters (Blue: F090W, Cyan: F150W, Yellow: F250M, Red: F430M). The one shown in \ref{fig:conv_simp} is WLM, which is located in the dwarf galaxy in the Cetus constellation. It is approximately 3 billion light-years away from us, and the distance covered by the entire image is about 1,700 light-years. Therefore, for the NIRCam detector, we used a total of 24 images, with each image considering only one NIRCam channel. We adjusted the 24 images obtained from WEBB to a size of $2048 \times 2048$, thus creating a training set consisting of $24$ samples, with a total of $24$, $2048\times 2048$ datasets.

We train our network using the Adam optimizer~\citep{Kingma2014AdamAM}, and set $\beta1 = 0.9$ and $\beta2 = 0.999$. We trained for 20000 epochs using a piecewise constant decay learning rate. The specific learning rate decay is set as follows: $\mathbf{boundaries} = [1000, 2000, 4000, 8000, 14000]$; $\mathbf{values} = [0.1, 0.01, 0.001, 0.0005, 0.0001, 0.00005]$. This means that the learning rate is $0.1$ for epochs $0\sim 1000$, $0.01$ for epochs $1000\sim 2000$, and so on. We conducted all the experiments using the TensorFlow2 on NVIDIA A40.

\subsection{Results and Discussion}\label{sec:result_discussion}

\begin{table}[]
	\centering\caption{The initial column in this study delineates the diverse methods under comparison. The second column delineates the image quality matrix. The subsequent columns, from the third to the seventh, exhibit the Structural Similarity Index (SSIM) and Peak Signal-to-Noise Ratio (PSNR) results of images acquired through the aforementioned distinct methods. The concluding two methods involve our model utilizing Autoencoder (AE) and U-Net networks via the PI-AstroDeconv method.}
	\begin{tabular}{cccccccc}
		
		\hline
		\multirow{1}{*}{\bf{Methods}} & {\bf{Image Quality Metrics}} &\multicolumn{5}{c}{\bf{Images}} \\
		\cmidrule(rl){3-7}
		&  & {\bf{No. 1}} &{\bf{No. 2}} &  {\bf{No. 3}} &{\bf{No. 4}} &  {\bf{No. 5}} \\
		
		\hline
		\hline
		
		\multirow{2}{*}{\citet{treitel1969predictive}} 	        & SSIM & 0.6688   & 0.5452   & 0.6683   & 0.6611 &   0.5903      \\
		& PSNR(dB) & 23.16   & 21.13   & 25.97   & 20.85 &  21.67       \\
		
		\midrule
		\multirow{2}{*}{\citet{Bai8488519}}	        & SSIM & 0.7305   & 0.3581   & 0.7298   & 0.7033 &   0.6720       \\
		& PSNR(dB) & 19.13   & 20.05   & 24.90   & 19.71 &  22.04       \\
		
		\midrule
		\multirow{2}{*}{\citet{Chen9578235}}	& SSIM & \bf{0.7842}   & 0.5967   & 0.6463   & 0.8085 &   \bf{0.7674}       \\
		& PSNR(dB) & 25.24   & 23.78   & 25.69   & 25.80 &  \bf{27.31}      \\
		
		\midrule
		\multirow{2}{*}{\citet{Nan9157433}}	& SSIM & 0.7752   & 0.5686   & 0.6129   & 0.7127 &   0.6406       \\
		& PSNR(dB) & 24.30   & 22.41   & 24.75   & 22.35 &  23.54       \\
		
		\midrule
		\multirow{2}{*}{\citet{Ren9156633}}	    & SSIM & 0.6683   & 0.7688   & 0.7173   & 0.5368 &   0.5837      \\
		& PSNR(dB) & 24.59   & 25.99   & 25.94   & 23.21 &  25.21       \\

		\midrule
		\multirow{2}{*}{\makecell[c]{PI-AstroDeconv\\(AutoEncoder)}}	& SSIM & 0.7373   & 0.8116   & 0.7356  & 0.7072 &   0.6960       \\
		& PSNR(dB) & 24.28   & 29.81   & 25.06   & 25.94 &  19.60      \\
		
		\midrule
		\multirow{2}{*}{\makecell[c]{PI-AstroDeconv\\(U-Net)}}	& SSIM & 0.7566   & \bf{0.8368}   & \bf{0.7391}    & \bf{0.8170} &  0.7407       \\
		& PSNR(dB) & \bf{28.27}   & \bf{32.30}   & \bf{25.81}   & \bf{29.61} &   25.02      \\                           
		\bottomrule
	\end{tabular}
	\label{title:result}
\end{table}

In \ref{fig:conv_simp}, the simulation formula for generating input data is presented. The image on the left corresponds to the pristine image, which serves as our primary objective. The middle image showcases the PSF or beam of the telescope. The image on the right illustrates the effect of blurring by the PSF or beam. The accompanying equation in the figure explicitly demonstrates the convolution of the pristine image with the PSF, resulting in the creation of a blurred image. Both the PSF and blurred images are essential components of the required data for the PI-AstroDeconv network.

In \ref{fig:image_patch}, we showcase the input, output, and deconvolved images generated by our model. Owing to the inherent design of our network architecture, the input and output images are identical. Consequently, the image on the left serves as a visual representation of both the input and output of the PI-AstroDeconv network architecture, while the image on the middle and right provide a visual representation of the deconvolution results achieved by the network. As shown in \ref{fig:image_patch}, PI-AstroDeconv is able to noticeably enhance the quality of the deblurred images. The relatively faint and weak galaxies, marked by yellow boxes, exhibit greater prominence in the restored star chart than in the observed images. Moreover, the restoration of the WLM dwarf galaxy, indicated by the red box in \ref{fig:image_patch}, successfully eliminates blurring effects. However, it does not fully restore the eruptive galaxy, situated in the bottom right of the image to the right in \ref{fig:image_patch}, to its original linear state, leading to comparatively inferior outcomes.

Furthermore, a comprehensive analysis of the results was conducted. The performance of five different methods was compared to our two approaches using the same set of five images as presented in \ref{title:result}. The results were quantified using image quality matrices SSIM and PSNR. Based on these results, our method exhibits distinct advantages, attaining optimal results in four out of the five images. Some of the other methods exhibited subpar performance primarily due to their complete lack of prior information. Moreover, the image blurring effects induced by convolution differs significantly from conventional blurring effects. This distinction arises due to the fact that the Point Spread Function (PSF) affects almost every pixel in the image.


\section{Conclusion}\label{sec:conclusion}
Despite some progress made by traditional algorithms, such as regularized filter, Wiener filter, and Lucy-Richardson method, their effectiveness in deconvolution is not desirable. Developing unsupervised algorithms to eliminate beam effects is a challenging task and an active research area in astronomical data processing. It is necessary to design an unsupervised learning architecture, such as an AE, to capture the underlying structure and features of beam or PSF. These architectures can reveal the complex relationship between clean and anomalous images without explicit labels or annotations. Reconstructing clean images can effectively minimize the impact of anomalous beams and uncover the underlying cosmic signals.

Reconstructing clean images can effectively reduce the impact of aberrant beams and reveal potential cosmological signals. In this study, we introduce an unsupervised deep learning technique that incorporates physical prior information, specifically designed to address the challenge of blind image deconvolution. We propose the PI-AstroDeconv architecture that can be applied to various conventional deep learning models to perform deconvolution operations through training. The design of this architecture also allows for the use of multiple PSF or beam to address the issue of inaccurate PSF or beam measurements, which is one of our future research directions. In the last layer of the regression network, we incorporate the telescope beam or PSF, carefully setting the input and output to be the same image, in order to achieve the goal of deconvolution. Despite the existence of multiple potential solutions for deconvolution, our approach preserves the overall contour of the image under the guidance of neural network training. With advanced deep learning methods, our network aims to generate images that are very close to the original image to ensure the accuracy and reliability of the results.

In our future research, we will strive for continuous enhancements to our model, including the exploration of alternative networks such as Vision Transformer, in order to achieve superior outcomes. Furthermore, our plans encompass the application of this model to various telescopes, such as the Five-hundred-meter Aperture Spherical Radio Telescope~(FAST)~\citep{Li8331324}, the Square Kilometre Array~(SKA)~\citep{Dewdney5136190}, and the upcoming the China Space Station Telescope~(CSST)~\citep{zhan2018overview}, with the objective of acquiring enhanced image quality and illuminating a broader range of astronomical phenomena.

\section{Acknowledgments}
This research was partially supported by National Key R\&D Program of China (No. 2022YFB4501405), and Zhejiang Provincial Natural Science Foundation of China under Grant No. LY24A030001.

\bibliographystyle{unsrtnat}
\bibliography{mybibfile}

\end{document}